\documentclass[useAMS]{mn2e}

\usepackage{latexsym,graphicx}
\usepackage{color}
\usepackage{float}

\usepackage{setspace}

\newcommand\kms{{\rm\,km\,s^{-1}}}
\newcommand\msun{\rm\,M_\odot}

\newcommand\myr{\msun \, {\rm yr}^{-1}}

\def\apgt{\ {\raise-.5ex\hbox{$\buildrel>\over\sim$}}\ }
\def\aplt{\ {\raise-.5ex\hbox{$\buildrel<\over\sim$}}\ }

\title[Stability of bow shocks generated by RSGs]{On the stability of bow shocks generated by red supergiants: the case of IRC\,$-$10414}

\author[D. M.-A.~Meyer et al.]
       {D. M.-A.~Meyer,$^{1}$\thanks{E-mail: dmeyer@astro.uni-bonn.de}
       V. V.~Gvaramadze,$^{2,3}$ N.~Langer,$^{1}$ J.~Mackey,$^{1}$ P.~Boumis,$^4$
       \newauthor
       and S.~Mohamed $^5$\\
        $^{1}$Argelander-Institut f\"ur Astronomie der Universit\"at Bonn, Auf dem H\"ugel 71, 53121, Bonn, Germany \\
        $^{2}$Sternberg Astronomical Institute, Lomonosov Moscow State University, Universitetskij Pr. 13, Moscow 119992, Russia\\
        $^{3}$Isaac Newton Institute of Chile, Moscow Branch, Universitetskij Pr. 13, Moscow 119992, Russia\\
        $^{4}$Institute for Astronomy, Astrophysics, Space Applications and Remote Sensing, National Observatory
        of Athens, \\ I. Metaxa \& Vas. Pavlou St., Palaia Penteli, 15236 Athens, Greece\\
        $^{5}$South African Astronomical Observatory, Observatory Road, Observatory, Cape Town, 7925, South Africa
        }
\begin{document}

\date{Accepted 2013 December 05. Received 2013 December 05; in original form 2013 November 18}

\maketitle

\label{firstpage}

\begin{abstract}
In this Letter, we explore the hypothesis that the smooth
appearance of bow shocks around some red supergiants (RSGs) might
be caused by the ionization of their winds by external sources of
radiation. Our numerical simulations of the bow shock generated by
IRC\,$-$10414 (the first-ever RSG with an optically detected bow
shock) show that the ionization of the wind results in its
acceleration by a factor of two, which reduces the difference
between the wind and space velocities of the star and makes the
contact discontinuity of the bow shock stable for a range of
stellar space velocities and mass-loss rates. Our best fit model
reproduces the overall shape and surface brightness of the
observed bow shock and suggests that the space velocity and
mass-loss rate of IRC\,$-$10414 are $\approx$50 $\kms$ and
$\approx$$10^{-6}$ $\myr$, respectively, and that the number
density of the local ISM is $\approx$3 ${\rm cm}^{-3}$. It also
shows that the bow shock emission comes mainly from the shocked
stellar wind. This naturally explains the enhanced nitrogen
abundance in the line-emitting material, derived from the
spectroscopy of the bow shock. We found that photoionized bow
shocks are $\approx$15$-$50 times brighter in optical line
emission than their neutral counterparts, from which we conclude
that the bow shock of IRC\,$-$10414 must be photoionized.
\end{abstract}


\section{Introduction}
\label{sec:intro}

\begin{figure*}
\includegraphics[width=14.0cm]{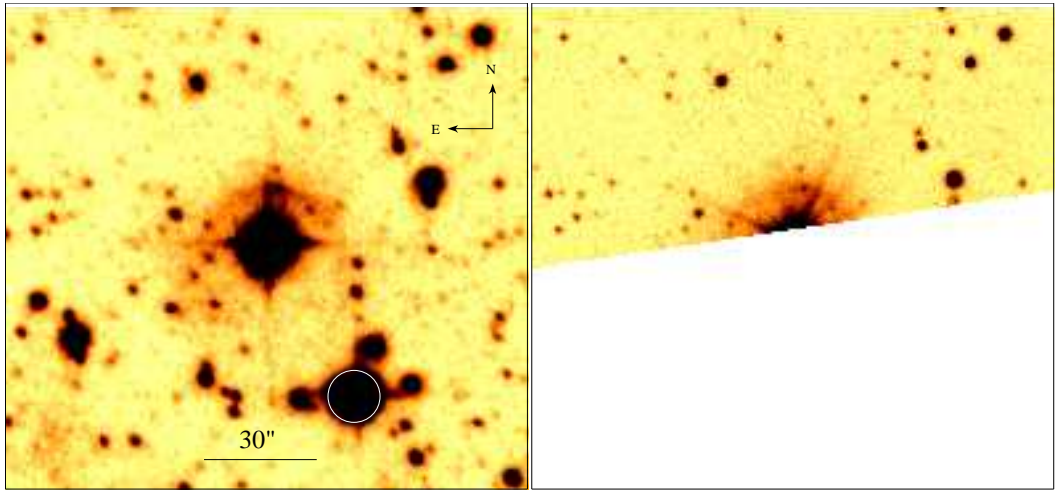}
\centering \caption{Left: H$\alpha+$[N{\sc ii}] image of
IRC\,$-$10414 and its bow shock from the SHS. The WC5 star
WR\,114, located at $\approx$45 arcsec southwest of IRC\,$-$10414,
is marked by a white circle. Right: H$\alpha+$[N{\sc ii}] image of
the bow shock obtained with the 2.3-m Aristarchos telescope (the
image is truncated because the CCD detector was offset from
IRC\,$-$10414 to avoid saturation). The orientation and the scale
of the images are the same. See the text for details. At a
distance of 2 kpc, 30 arcsec correspond to $\approx$0.29 pc.}
\label{fig:bow}
\end{figure*}

A significant fraction of runaway OB stars are moving
supersonically through the local interstellar medium (ISM)
(Huthoff \& Kaper 2002) and therefore generate bow shocks. The
detection of these arc-like structures serves as an indication
that their associated stars are massive enough to possess strong
winds and could be used for i) identifying distant and/or highly
reddened (runaway) OB stars (e.g. Gvaramadze, Kroupa \&
Pflamm-Altenburg 2010), ii) searching for parent clusters to these
stars (e.g. Gvaramadze \& Bomans 2008) and iii) constraining their
mass-loss rates (Kobulnicky, Gilbert \& Kiminki 2010; Gvaramadze,
Langer \& Mackey 2012) and parameters of the local ISM (Kaper et
al. 1997; Gvaramadze et al. 2013, hereafter Paper\,I).

Analytical (Dgani, van Buren \& Noriega-Crespo 1996) and numerical
studies (Comer\'{o}n \& Kaper 1998; Blondin \& Koerwer 1998) of
bow shocks show that they are subject to different kinds of
instabilities, which along with density inhomogeneities and
interstellar magnetic field can significantly affect their
appearance. In particular, Dgani et al. (1996) showed that
isothermal bow shocks are unstable if the stellar space velocity,
$v_*$, is larger than the wind velocity, $v_{\rm w}$. This
condition is usually fulfilled by cool runaway stars, e.g., red
supergiants (RSGs), whose wind velocities of $\approx$$20 \, \kms$
are comparable to or less than their typical space velocities of
several tens of $\kms$. Numerical simulations of bow shocks
produced by RSGs (Brighenti \& D'Ercole 1995; Mohamed, Mackey \&
Langer 2012; Cox et al. 2012; Decin et al. 2012) confirmed that
they are indeed generally unstable to a significant degree. This
result is in conflict with the observed smoothness of bow shocks
associated with two of the three known bow-shock-producing RSGs,
namely Betelgeuse (Noriega-Crespo et al. 1997) and IRC\,$-$10414
(Paper\,I). On the other hand, a bow shock around the third of
these RSGs, $\mu$\,Cep, shows clear signatures of instabilities
(Cox et al. 2012) in good agreement with the theoretical and
numerical predictions. There should therefore exist some factors
which stabilize bow shocks around RSGs. Decin et al. (2012) began
to explore this subject by considering a bow shock from a RSG with
a neutral wind but moving in an ionized ISM (photoheated to
8000\,K), finding that this reduced the strength of instabilities
compared to a bow shock in a neutral ISM.

In this Letter, we use numerical simulations of the bow shock of
IRC\,$-$10414 to investigate the hypothesis that bow shocks
generated by RSGs could be stable if the stellar wind and the
ambient ISM are heated and ionized by an external source of
radiation. The relevant data on the bow shock are reviewed in
Section\,\ref{sec:irc}. The numerical models of the bow shock are
presented in Section\,\ref{sec:num} and discussed in
Section\,\ref{sec:dis}. We summarize in Section\,\ref{sec:sum}.

\begin{figure*}
\begin{minipage}{0.9\linewidth}
\begin{minipage}[h]{0.49\linewidth}
\center{\includegraphics[width=1\linewidth]{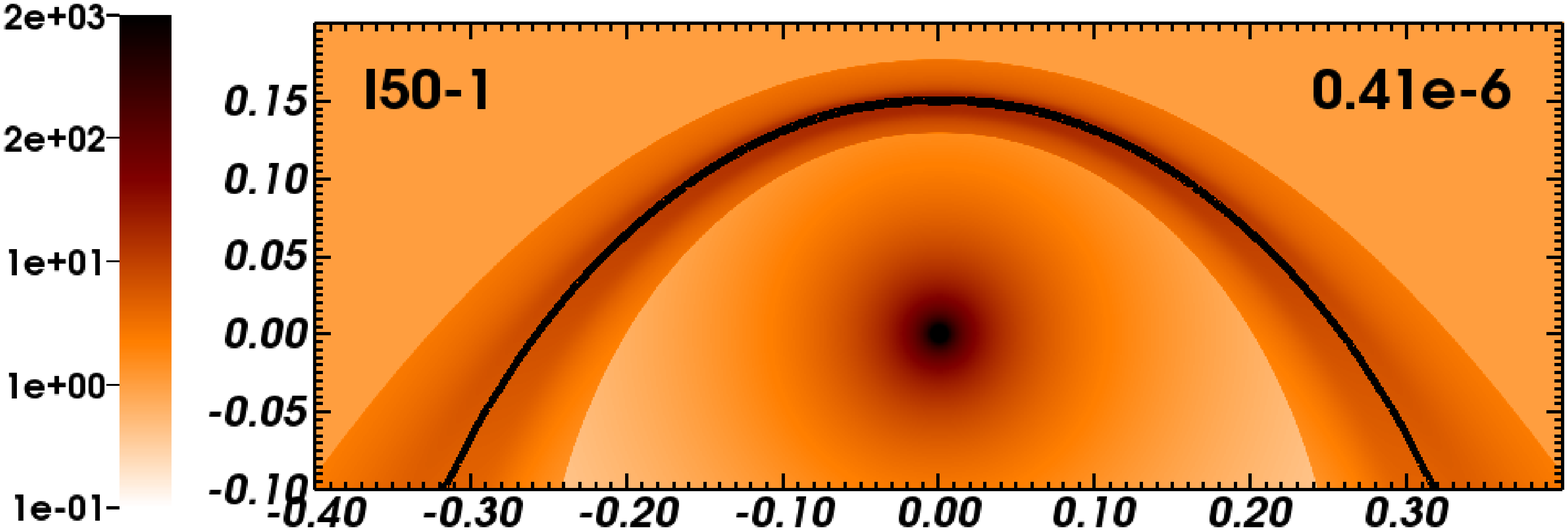}} {} \\
\end{minipage}
\hfill
\begin{minipage}[h]{0.49\linewidth}
\center{\includegraphics[width=1\linewidth]{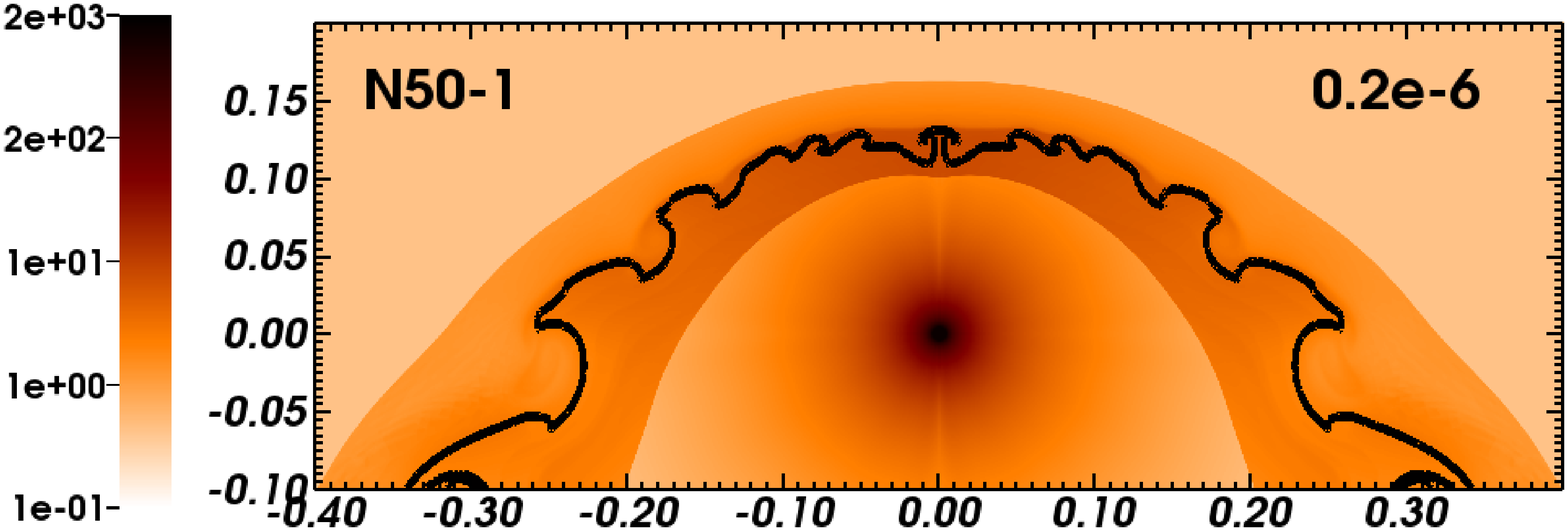}} {} \\
\end{minipage}
\vfill
\begin{minipage}[h]{0.49\linewidth}
\center{\includegraphics[width=1\linewidth]{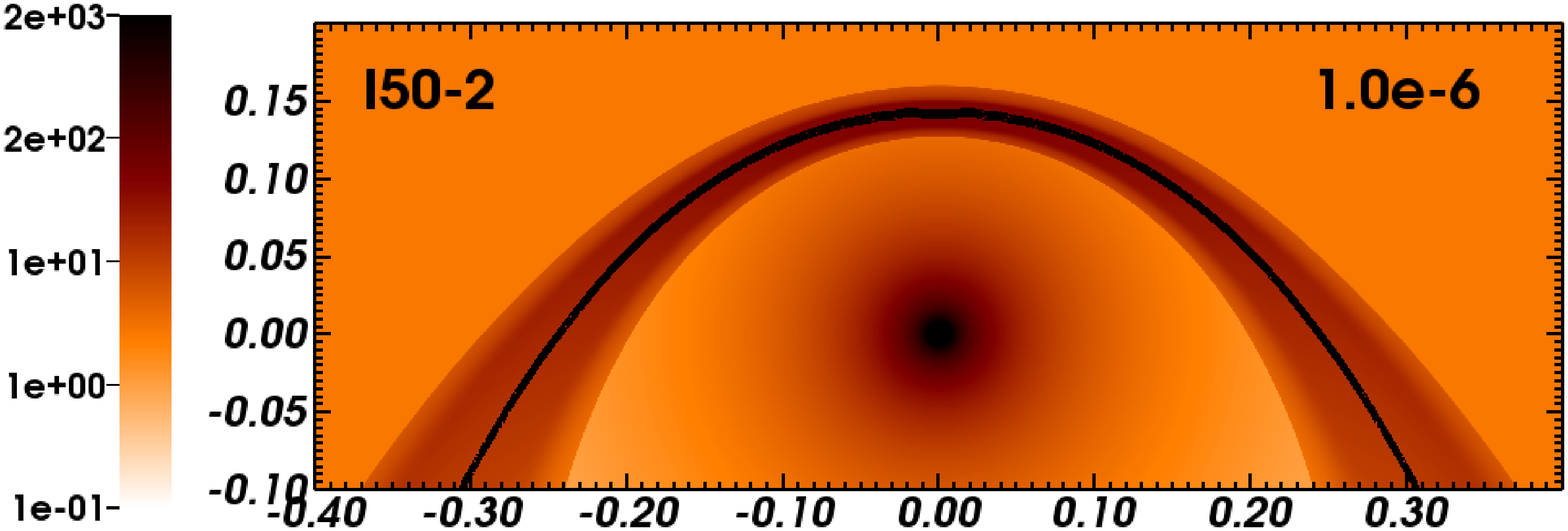}} {} \\
\end{minipage}
\hfill
\begin{minipage}[h]{0.49\linewidth}
\center{\includegraphics[width=1\linewidth]{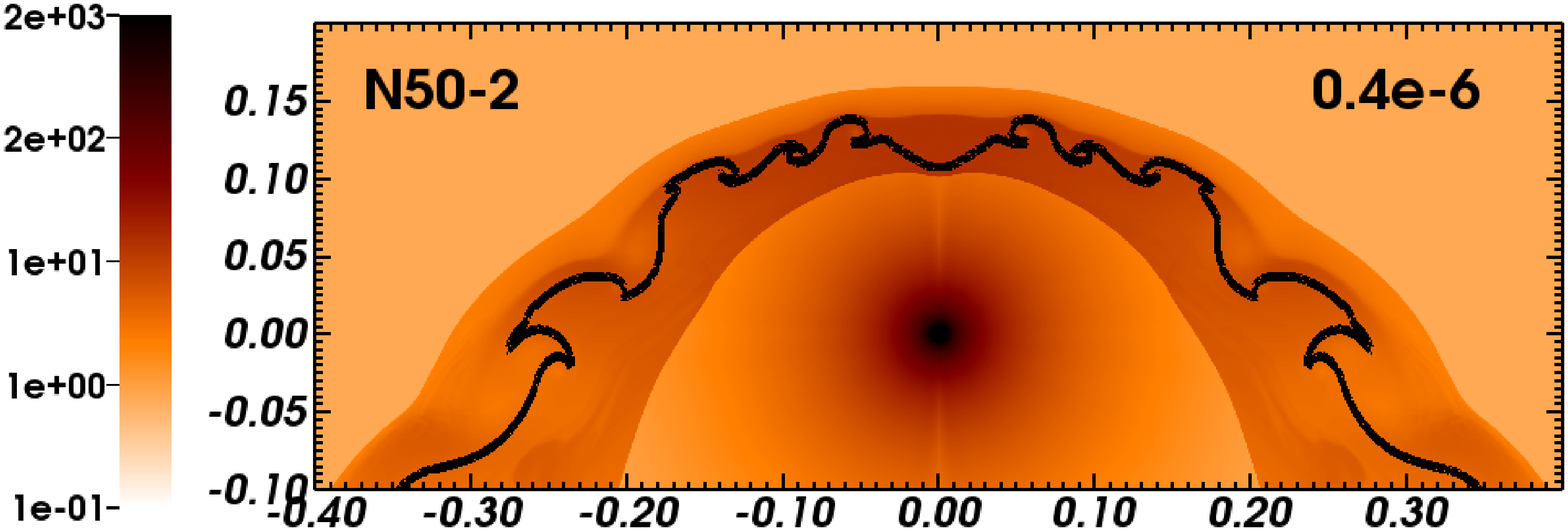}} {} \\
\end{minipage}
\hfill
\begin{minipage}[h]{0.49\linewidth}
\center{\includegraphics[width=1\linewidth]{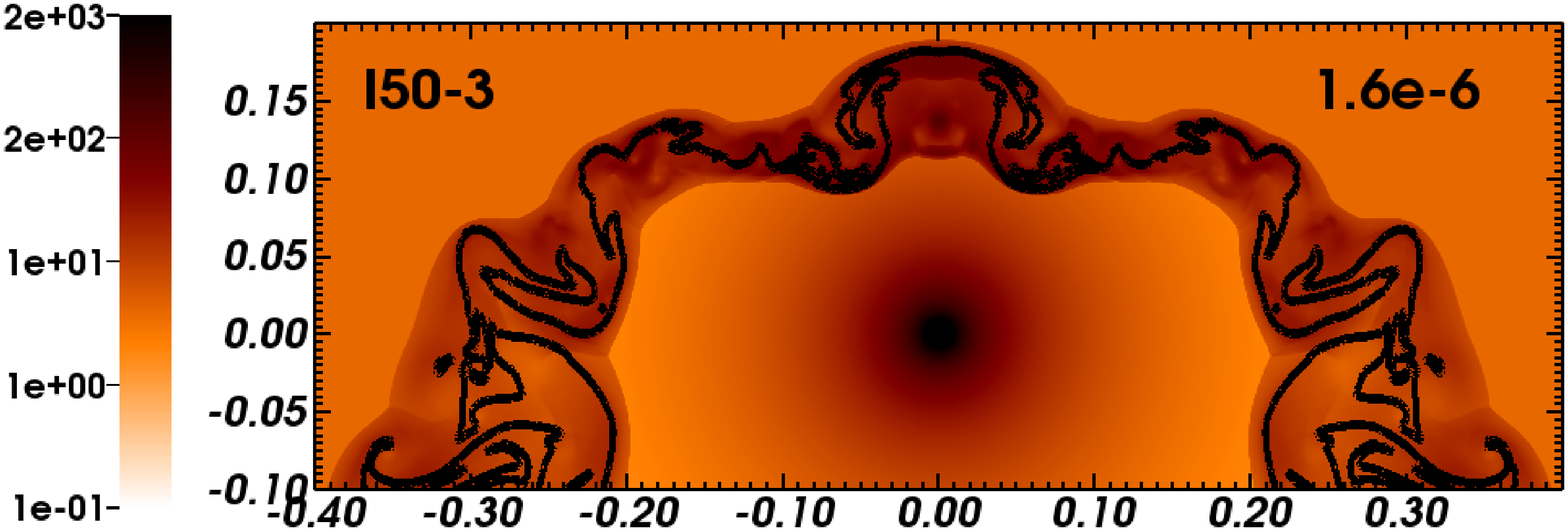}} {} \\
\end{minipage}
\hfill
\begin{minipage}[h]{0.49\linewidth}
\center{\includegraphics[width=1\linewidth]{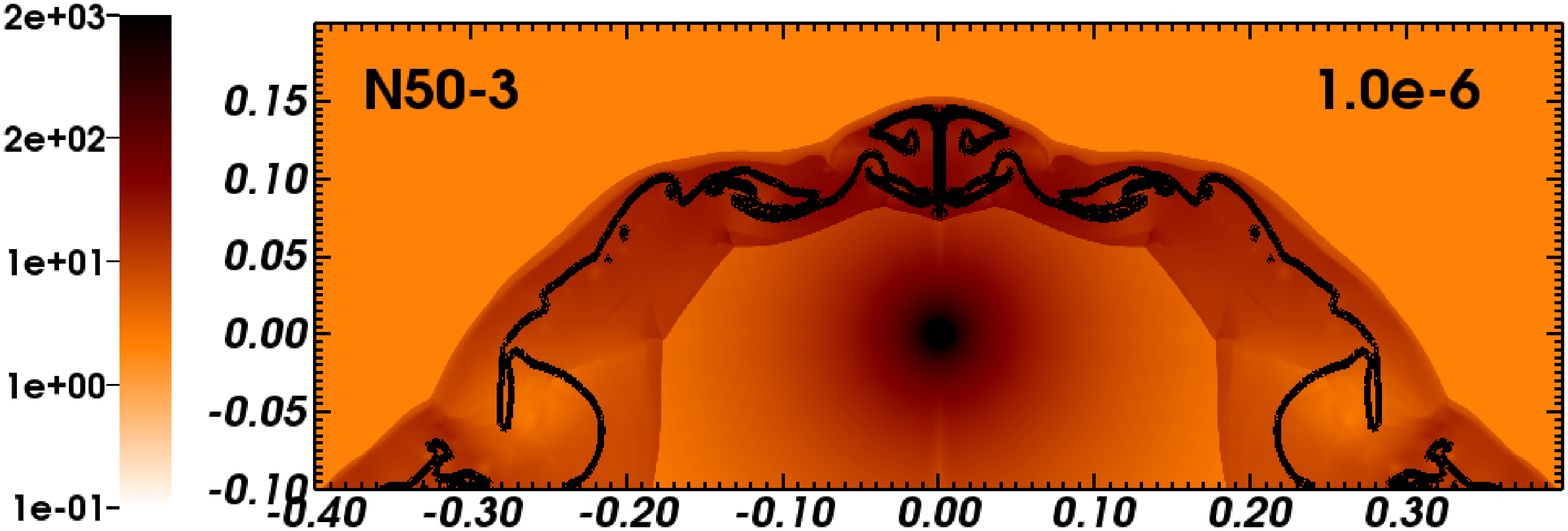}} {} \\
\end{minipage}
\end{minipage}
\caption{Grid of models of the bow shock generated by a RSG moving
at $50 \, \kms$. The panels show the gas number density plotted on
the logarithmic scale in units of ${\rm cm}^{-3}$. The left-hand
panels present the fully ionized models and the right-hand panels
the neutral ones. On each panel the left-hand key refers to the
nomenclature detailed in Table\,\ref{tab:par} and the right-hand
key indicates $\dot{M}$ in units of $\myr$. The solid line traces
the position of the contact discontinuity. Models are shown at
least $0.1\, \rm Myr$ after the beginning of the simulations. The
$x$-axis corresponds to the radial direction and the symmetry axis
is aligned with the space velocity of the star (both axes are in
units of pc). Note that not all of the computational domain is
shown.} \label{fig:I50}
\end{figure*}

\section{The bow shock of IRC\,$-$10414}
\label{sec:irc}

The bow shock around IRC\,$-$10414 is the first-ever optically
detected bow shock generated by RSGs (the data on the bow shock
and IRC\,$-$10414 quoted below are from Paper\,I unless otherwise
stated). The left-hand panel of Fig.\,\ref{fig:bow} presents the
discovery H$\alpha$+[N\,{\sc ii}] $\lambda\lambda$6548, 6584 image
of the bow shock from the SuperCOSMOS H-alpha Survey (SHS; Parker
et al. 2005), showing a smooth arc-like nebula at $\approx$15
arcsec from the star. At a distance to IRC\,$-$10414 of 2 kpc
(Maeda et al. 2008), the stand-off distance of the bow shock is
$R_{\rm SO}\approx$0.14 pc. For an inclination angle of the bow
shock to the plane of the sky of $\approx$$20\degr$, the
projection effect on the observed $R_{\rm SO}$ is negligible (see
Gvaramadze et al. 2011). The right-hand panel of
Fig.\,\ref{fig:bow} shows a follow-up image of the bow shock
obtained with the 2.3-m Aristarchos f/8 telescope at Helmos
Observatory, Greece on 2013 August 9 with 1800 s exposure, through
a 40 \AA \, bandwidth filter centred on the H$\alpha$+[N\,{\sc
ii}] lines. To avoid saturation, IRC\,$-$10414 was placed outside
the CCD detector. Although the resolution of this image is about
three times higher than that of the SHS one, the bow shock still
does not show any signatures of instabilities.

Using equation\,(1) and the flux calibration factor of 16.1 counts
pixel$^{-1}$ R$^{-1}$ from Table\,1 in Frew et al. (2013), we
derived from the SHS image the surface brightness of the bow shock
near the apex of $\Sigma _{\rm obs}$$\approx$77 R
(1R$\equiv$1Rayleigh=5.66$\times$$10^{-18}$ erg cm$^{-2}$ s$^{-1}$
arcsec$^{-2}$ at H$\alpha$).

Optical spectroscopy of the bow shock (carried out with the
Southern African Large Telescope) showed that the line-emitting
material is enriched in nitrogen, which implies that the emission
at least partially originates from the shocked stellar wind (cf.
Section\,\ref{sec:dis}). It also allowed us to constrain the
number density of the ambient ISM to be $n_0$$\leq$5 ${\rm
cm}^{-3} \, (v_*/70 \, \kms)^{-2}$ (see Paper\,I for details),
which in its turn imposes a limitation on the mass-loss rate,
$\dot{M}$, of IRC\,$-$10414. The space velocity of IRC\,$-$10414
$v_*$$\approx$70$\pm$20\,$ \kms$ is several times higher than the
stellar wind velocity $v_{\rm w}$=21$\pm$$2\,\kms$, derived from
maser observations of a region within a few hundreds of AU from
the star. Thus, if the bow shock of IRC\,$-$10414 is a thin shell
(a natural assumption for bow shocks produced by cool stars),
then, according to Dgani et al. (1996), it should be unstable and
ragged, which clearly contradicts our observations.

Previous numerical simulations of bow shocks generated by RSGs
proceeded from the common assumption that the stellar wind is
neutral. This assumption, however, would be invalidated if the
wind is ionized by an external source of radiation, like a nearby
hot massive star or a star cluster. Good examples of such
situation are ionized nebulae around the RSGs NML\,Cyg and W26,
which are the result of ionization of the stellar wind by the
nearby association Cyg\,OB2 (Morris \& Jura 1983) and star cluster
Westerlund\,1 (Wright et al. 2013), respectively. As discussed in
Paper\,I, the ionization of the wind might exert a stabilizing
influence on RSG bow shocks. Proceeding from this, we proposed
that the smooth shape of the bow shock around IRC\,$-$10414 is
because the wind of this star and the ambient ISM are ionized by
the nearby WC5 star WR\,114 (see Fig.\,\ref{fig:bow}) and/or by
the massive star cluster NGC\,6611. In this connection, we note
that the spectrum of the bow shock shows very strong [N\,{\sc ii}]
$\lambda\lambda$6548, 6583 emission lines (see fig.\,3 in
Paper\,I), which means that the stellar wind is ionized to a
significant degree (cf. Section\,\ref{sec:dis}). Since the wind
material cannot be collisionally ionized because the reverse shock
is too weak, it is natural to assume that the wind is photoionized
by an external source. The above considerations motivated us to
carry out numerical simulations presented in this Letter.

\begin{table}
\centering{ \caption{Input parameters of the grid models and the
surface brightness of the model bow shocks. Columns show,
respectively, model identifier, space velocity in $\kms$,
mass-loss rate in $10^{-6} \, \myr$, ambient ISM number density in
${\rm cm}^{-3}$, and maximum surface brightness of the models,
before and after correction for the interstellar extinction
towards IRC\,$-$10414, in Rayleighs.} \label{tab:par}
\begin{tabular}{lccccccr} \hline
Model & $v_*$ & $\dot{M}$ & $n_0$ & $\Sigma _{\rm max}$ & $\Sigma _{\rm max} ^{\rm cor}$ \\
\hline
I50-1 & 50 & 0.41  & 1.21  & 133.2 & 5.3 \\
I50-2 & 50 & 1.01  & 3.30  & 1176.7 & 47.1 \\
I50-3 & 50 & 1.62  & 5.00  & 8436.8 & 337.5 \\
I70-1 & 70 & 0.41  & 0.73  & 155.4 & 6.2 \\
I70-2 & 70 & 1.01  & 1.51  & 865.9 & 34.6 \\
I90-1 & 90 & 0.41  & 0.41  & 144.3 & 5.8 \\
\hline
N50-1 & 50 & 0.20  & 0.48 & 3.1& 0.1 \\
N50-2 & 50 & 0.41  & 0.91 & 8.0 & 0.3 \\
N50-3 & 50 & 1.01  & 2.60 & 33.3 & 1.3 \\
N70-1 & 70 & 0.41  & 0.35 & 3.1 & 0.1 \\
N90-1 & 90 & 0.41  & 0.26 & 9.5 & 0.4 \\
\hline
\end{tabular}
}
\end{table}

\section{Numerical simulations}
\label{sec:num}

We performed 2D numerical simulations using the {\sc pluto} code
(Mignone et al. 2007, 2012). The simulations were carried out in
cylindrical coordinates on a uniform grid of size of
$[0,0.4]\times[-0.1,0.3]$ pc and spatial resolution of $2.25\times
10^{-4}\, \rm{pc}\, \rm{cell}^{-1}$. A stellar wind was injected
into the computational domain via an half circle of radius of 20
cells ($\approx$900 AU) centred at the origin, and its interaction
with the ISM was modelled in the reference frame of the star. Wind
material is distinguished from the ISM using a passive tracer
advected together with the fluid. The ISM composition is assumed
to be solar (Asplund et al. 2009).

Optically-thin radiative cooling and heating were taken into
account. For a fully ionized medium, the cooling curve is the sum
of contributions from H, He and metals (Wiersma, Schaye \& Smith
2009), collisionally excited forbidden lines (Henney et al. 2009)
and H recombination together with heating from the reionization of
recombining H atoms (Osterbrock \& Bochkarev 1989; Hummer 1994).
The equilibrium temperature of this curve is $\approx$8000 K. For
models with the neutral medium, the cooling curve is the sum of
contributions from H, He and metals (Wiersma et al. 2009) and the
dust heating by the Galactic far-ultraviolet background (Wolfire
et al. 2003). The equilibrium temperature of this curve is
$\approx$3300 K for $n_0$=$1 \, {\rm cm}^{-3}$. All models include
electronic thermal conduction (Cowie \& McKee 1977).

We have run a grid of 11 models, in which both the stellar wind
and the ISM were considered to be either fully ionized or neutral,
labelled `I' and `N', respectively. Three space velocities were
considered, $v_* =50, 70$ and $90 \, \kms$, and $v_{\rm w}=21 \,
\kms$ was set in all models. For $\dot{M}$ we adopted a range of
values based on various mass-loss prescriptions proposed for RSGs
(see Paper\,I), ranging from $\approx$0.4$\times 10^{-6}$ to
1.6$\times 10^{-6} \, \myr$. For each model, $n_0$ was adjusted in
such a way that $R_{\rm SO}$ of a model bow shock is equal to the
observed one, when a steady state was reached. All models are run
for at least 0.1 Myr, which corresponds to more than 16 grid
crossing times. The parameters of the models are summarized in
Table\,\ref{tab:par}.

For the sake of comparison with observations, we calculated the
H$\alpha+$[N{\sc ii}] surface brightness of the model bow shocks
using the prescriptions by Dopita (1973) and Osterbrock \&
Bochkarev (1989). To reproduce the observed [N\,{\sc
ii}]/H$\alpha$ line ratio of 2.4$\pm$0.1 (measured in the spectrum
of the bow shock at an angle of 25$\degr$ from the apex; see
Paper\,I), we assumed that the RSG wind is enriched in nitrogen by
a factor of 6 (cf. Brott et al. 2011). The maximum value of the
brightness, $\Sigma _{\rm max}$, for each model is given in the
column 5 of Table\,\ref{tab:par}, while the column 6 gives $\Sigma
_{\rm max}$ corrected for the interstellar extinction towards
IRC\,$-$10414, which in the $R$-band is $\approx$3.5 mag
(Paper\,I).

\section{Results and discussion}
\label{sec:dis}

\begin{figure}
\includegraphics[width=8.5cm,angle=0]{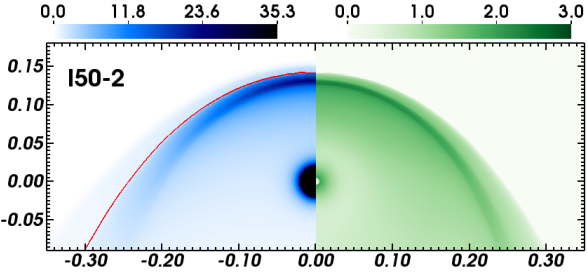}
\centering \caption{Left: Surface brightness for the ionized model
I50-2 plotted on a linear scale in units of R. The solid (red)
line traces the position of the contact discontinuity. Right:
[N{\sc ii}]/H$\alpha$ line ratio for the same model.}
\label{fig:emi}
\end{figure}
\begin{figure}
\center{\includegraphics[width=8.5cm,angle=0]{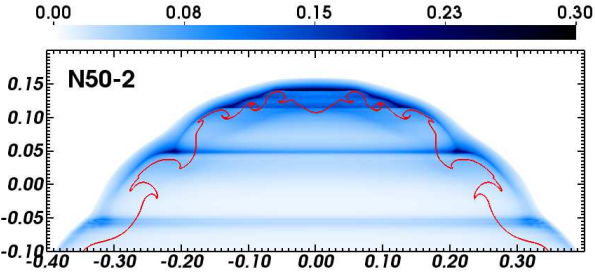}}
\centering \caption{Surface brightness for the neutral model N50-2
plotted on a linear scale in units of R. The solid (red) line
traces the position of the contact discontinuity} \label{fig:emn}
\end{figure}
\begin{figure}
\includegraphics[width=0.6\linewidth,angle=270]{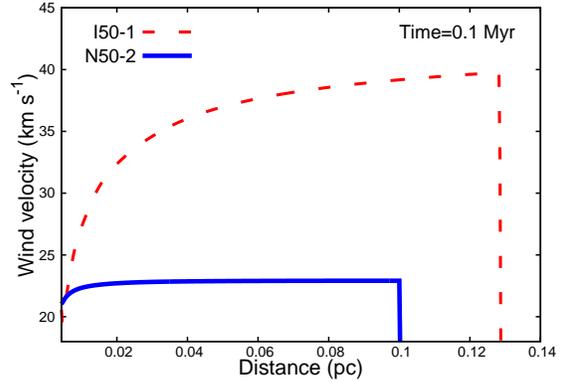}
\centering \caption{Wind velocity profiles along the $y$-axis for
the models I50-1 and N50-2. Heating of the wind by photoionization
results in its acceleration by a factor of two (see the text for
details).} \label{fig:vel}
\end{figure}

Fig.\,\ref{fig:I50} plots the gas number density of the ionized
(left-hand panels) and neutral bow shocks generated by RSGs moving
with a velocity of 50 $\kms$. The solid (black) line traces the
position of the contact discontinuity. One can see that the
ionized bow shocks are stable as long as
$\dot{M}$$\la$1$\times$$10^{-6}$ $\myr$. For higher $\dot{M}$
and/or $v_*$, the ionized bow shocks became unstable (e.g. model
I50-3 in Fig.\,\ref{fig:I50}), which prevent them from reaching a
steady state. Of three neutral models shown in Fig.\,\ref{fig:I50}
two ones have smooth forward and reverse shocks, while the contact
discontinuity in all of them is very ragged. For
$\dot{M}$$\ga$$10^{-6}$ $\myr$ and/or $v_*$$>$50 $\kms$, the
contact discontinuity becomes even more unstable and the overall
shape of the neutral bow shocks becomes highly distorted (e.g.
model N50-3 in Fig.\,\ref{fig:I50}). We therefore expect that one
of the models I50-1, I50-2, N50-1 and N50-2 could represent the
bow shock around IRC\,$-$10414.

To substantiate this expectation, we compare $\Sigma_{\rm obs}$
with the model predictions. An inspection of Table\,\ref{tab:par}
shows that the higher $\dot{M}$ the higher $\Sigma _{\rm max}$ of
the models. Three of the ionized models have $\Sigma _{\rm max}
^{\rm cor}$ comparable to or larger than $\Sigma _{\rm obs}$. Of
these models only I50-2 is stable (see Fig.\,\ref{fig:I50}).
$\Sigma _{\rm max} ^{\rm cor}$ of the remaining three ionized
models slightly exceeds the sensitivity limit to diffuse emission
of the SHS of 2$-$5 R (Parker et al. 2005) and therefore, in
principle, these bow shocks could be detected with this survey. On
the contrary, all the neutral models are so dim that their $\Sigma
_{\rm max} ^{\rm cor}$ is below the sensitivity limit of the SHS.
Thus, we conclude that I50-2 is the best fit model of the bow
shock of IRC\,$-$10414. Interestingly, $\dot{M}$ adopted in this
model is a factor of 5$-$10 smaller than that predicted by most of
the mass-loss prescriptions proposed for RSGs (see Paper\,I) and a
factor of two higher than what follows from the recipe by
Verhoelst et al. (2009).

The left-hand panel of Fig.\,\ref{fig:emi} presents a map of the
surface brightness in the H$\alpha+$[N{\sc ii}] lines for our
preferred model I50-2. It shows that the emission comes mainly
from the shocked wind, which naturally explains why the brightness
of the ionized models increases with $\dot{M}$ (see
Table\,\ref{tab:par}) and why the line-emitting material in the
bow shock of IRC\,$-$10414 is enriched in nitrogen. The right-hand
panel of Fig.\,\ref{fig:emi} plots the [N\,{\sc ii}]/H$\alpha$
line ratio for the emission originating from the shocked wind for
the same model. This ratio has a value of 2.3 for an angle of
25$\degr$ from the apex of the bow shock, which is in a good
agreement with the observed value of 2.4$\pm$0.1.

Fig.\,\ref{fig:emn} plots the surface brightness for the neutral
model N50-2. Unlike the ionized models, the H$\alpha+$[N{\sc ii}]
emission in the neutral ones originates from the collisionally
ionized ISM, i.e. just behind the forward shock. Correspondingly,
the model N50-2 has a rather smooth appearance, but its $\Sigma
_{\rm max} ^{\rm cor}$ is below the sensitivity limit of the SHS.
This further supports our claim that the bow shock of
IRC\,$-$10414 is ionized by an external source.

Using Fig.\,\ref{fig:emi}, we measured a ratio $R_{\rm
SO}/R(\theta)$, where $R(\theta)$ is the distance between the star
and the bow shock at an angle $\theta$ from the apex. For
$\theta\approx$$75\degr$ (the half-opening angle of the observed
bow shock), we found $R_{\rm SO}/R(\theta)=1.36$, which is in a
reasonable agreement with both the observed ratio of 1.33 and the
theoretical one of 1.44, derived from the thin-shell bow shock
model by Wilkin (1996).

It should be noted that the ionized models have higher ISM
densities than the neutral ones with the same $v_{\star}$ and
$\dot{M}$ (see Table\,\ref{tab:par}). Since $R_{\rm SO}$ is fixed
in all models, this difference implies that the wind velocity is
higher in the ionized models. Indeed, the instantaneous heating of
the wind material up to $\approx$8000 K results in increase of its
thermal pressure, which in its turn leads to the wind acceleration
(see Oort \& Spitzer 1955 for more details). Fig.\,\ref{fig:vel}
shows the wind velocity profiles along the $y$-axis for the models
I50-1 and N50-2. In I50-1 the wind is accelerated by a factor of
two and its velocity becomes comparable to $v_*$, while in N50-2
the wind velocity remains constant\footnote{A slight acceleration
of the wind in this model is because of a boundary effect.}, i.e.
a factor of $\approx$2.5 less than $v_*$. Correspondingly, the
shear produced by the relative motion of the shocked wind and
shocked ISM is stronger in the N50-2 model, which makes it more
prone to the development of the Kelvin-Helmholtz instability at
the contact discontinuity. The same is true for the models with
higher $v_*$ because the growth time of the instability is
inversely proportional to the magnitude of the shear, which in its
turn increases with $v_*$.

\section{Summary}
\label{sec:sum}

In this Letter, we presented a grid of models of bow shocks
produced by RSGs, in which both the stellar wind and the ISM were
considered to be either fully ionized or neutral. We investigated
whether the smooth appearance of the bow shock around the RSG
IRC\,$-$10414 might be caused by the ionization of the stellar
wind by an external source of radiation. We found that although
both kinds of models can have a smooth appearance in the
H$\alpha$+[N\,{\sc ii}] lines, only the ionized ones can
simultaneously reproduce the overall shape and the brightness of
this bow shock. Our simulations showed that the ionization of the
stellar wind results in its acceleration by a factor of two, which
reduces the shear at the contact discontinuity and makes the bow
shock stable for a range of stellar space velocities and mass-loss
rates. Our best fit model of the bow shock suggests that the space
velocity and mass-loss rate of IRC\,$-$10414 are $\approx$50
$\kms$ and $\approx$$10^{-6}$ $\myr$, respectively, and that the
number density of the local ISM is $\approx$3 ${\rm cm}^{-3}$. We
found also that in the ionized models the H$\alpha$+[N\,{\sc ii}]
emission originates mostly from the shocked RSG wind, which
naturally explains why the line-emitting material in the bow shock
of IRC\,$-$10414 is enriched in nitrogen. Our results suggest that
the ionization of the stellar wind might be responsible for the
smooth appearance of bow shocks generated by other RSGs, or
asymptotic giant branch stars.

\section{Acknowledgements}
We are grateful to D.J. Frew for providing us with data before
publication. JM acknowledges funding from the Alexander von
Humboldt Foundation and the Deutsche Forschungsgemeinschaft
priority program 1573, `Physics of the Interstellar Medium'.
The Aristarchos telescope is operated on Helmos Observatory by the
Institute of Astronomy, Astrophysics, Space Applications and
Remote Sensing of the National Observatory of Athens. All the
simulations were run on the JUROPA supercomputer at the
J{\"u}elich Supercomputing Centre.

\end{document}